\documentclass[twocolumn]{article}
\usepackage[hmarginratio=1:1,top=30 mm,columnsep=20pt,hmargin=2.5cm]{geometry}% http://ctan.org/pkg/geometry
\usepackage[font=it]{caption}% http://ctan.org/pkg/caption
\usepackage{lipsum,graphicx,float}% http://ctan.org/pkg/{multicol,lipsum,graphicx,float}
\usepackage{authblk}
\usepackage[nodisplayskipstretch]{setspace} 
\title{Leveraging Adiabatic Quantum Computation for Election Forecasting}
\author[1]{Maxwell Henderson}
\author[2]{John Novak}
\author[1]{Tristan Cook}
\affil[1]{QxBranch}
\affil[2]{Standard Cognition}
\date{}  
\begin{document}

\maketitle

\begin{abstract}
Accurate, reliable sampling from fully-connected graphs with arbitrary correlations is a difficult problem.  Such sampling requires knowledge of the probabilities of observing every possible state of a graph.  As graph size grows, the number of model states becomes intractably large and efficient computation requires full sampling be replaced with heuristics and algorithms that are only approximations of full sampling.  This work investigates the potential impact of adiabatic quantum computation for sampling purposes, building on recent successes training Boltzmann machines using a quantum device.  We investigate the use case of quantum computation to train Boltzmann machines for predicting the 2016 Presidential election.
\end{abstract}

%\begin{multicols}{2}

\section{Introduction} \label{introduction}

As the results of the 2016 US Presidential Election were finalized, it was clear that the majority of professional polling groups, many of whom had predicted the probability of a Clinton victory to be well over 90\%, were had significantly overestimated their predictions (\cite{independent, huffingtonpost, cnn})).  While it could be argued that the underlying models were correct and that the particular result was just a very rare event, post-mortem analyses have revealed flaws that led to large prediction biases.  According to multiple post-election analyses, it was concluded that a leading cause of error in the majority of election forecasting models was a lack of correlation between individual states predictions (\cite{reuters, wbur, quanta, economist}).  Uncorrelated models, though much simpler to build and train, cannot capture the more complex behavior of a fully-connected system.  To capture these higher-order relationships, a fully-connected graphical model would be ideal.  While these models are more powerful, practical roadblocks have prevented their widespread adoption due to difficulties in implementation using classical computation.  However, recent studies have shown that quantum computing is a competitive alternative when generating such networks (\cite{2016arXiv160606123D, PhysRevX.7.041052, Benedetti2016, arXiv:1510.06356, Benedetti2017}).

Quantum machine learning (QML) is a blossoming field.  As summarized in the comprehensive review of 
QML in \cite{biamonte2017quantum}, machine learning applications from support vector machines to principal component analysis are being reimagined on various quantum devices.  One of the most exciting research areas within QML is deep quantum learning, which focuses on the impact quantum devices and algorithms can have on classical deep neural networks (DNNs) and graphical models.  A particular class of DNNs is the Boltzmann machine (BM), which is an incredibly powerful fully-connected graphical model that can be trained to learn arbitrary probability distributions.  A downside of these networks is that BMs are incredibly costly to train, a fact that has limited their practical application.  This large computational training cost has drawn attention to the implementation of quantum computation to help train such networks.  BMs realized on quantum devices (particularly adiabatic quantum devices such as those produced by D-Wave Systems (\cite{PhysRevX.7.041052}) may possess inherent benefits compared to those realized on classical devices.  Research groups have realized various forms of BMs (fully-connected BM, restricted Boltzmann machines (RBMs), and Helmholtz machines) trained using quantum computation, and this research has shown quantum computation can be used to effectively train neural networks for image recognition tasks (\cite{PhysRevX.7.041052, arXiv:1510.06356, Benedetti2017}).

In this work, we will leverage the power of adiabatic quantum computation to efficiently train fully-connected BMs for the novel purpose of election modeling.  Additionally, we have systematically explored a number of the assumptions underlying the approach of using adiabatic quantum computers (AQC) to model BMs, and we have demonstrated that for most systems of interest (such as this one) the approach does appear to be valid.  We believe the methods proposed in this paper could bring an interesting new factor into the conversation of election forecasting at large, one in which quantum computation could play a future role.

\section{Methodology} \label{methodology}

\subsection{Modeling Boltzmann Machines with AQC} \label{modeling_boltzmann_machines}

In this work, we will be generating fully-connected BMs trained using a D-Wave 2X 1152 qubit quantum device using the general method described in \cite{PhysRevX.7.041052}.  While the methodology for training both RBMs and BMs using a D-Wave machine have been laid out in previous papers (\cite{PhysRevX.7.041052, arXiv:1510.06356}), we will briefly review the logic and methodology here.

 A BM is a fully-connected graph of ($N$) binary units (neurons).  These neurons can be either ``visible" (directly model some aspect of a data distribution) or ``hidden" (not tied to any particular aspect of the data distribution and used only for capturing features from the data distribution).  Each network has $2^N$ possible states, and the probability of sampling a particular state $\textbf{s} = (s_1, ... , s_N)$ from the model is 
\begin{equation} \label{eq1}
p(\textbf{s}) = \frac{e^{-E(\textbf{s})}}{Z},
\end{equation}
 wherein $Z$ it the well-known partition function and $E$ is an energy function defined as
 \begin{equation} \label{eq2}
E(\textbf{s}) = -\sum_{s_i \in \textbf{s}} b_i s_i  - \sum_{s_i, s_j \in \textbf{s}} W_{ij} s_i s_j ,
\end{equation}
wherein $b$ represent the linear ``bias" on each unit and $W$ represents the ``weight" of the coupling between two units ($b$ and $W$ will be referred to as our ``model parameters").  To properly train the network, we need to adjust the model parameters so that the model distribution produced by repeatedly sampling the model is as close as possible to the underlying data distribution; more precisely, we want to maximize the log-likelihood, $L$, of the data distribution.  To calculate the model parameters for maximizing $L$, we use the familiar gradient descent method and learning rate $\eta$ to get model parameter update equations
 \begin{equation} \label{eq3}
\Delta W_{ij} = \frac{1}{\eta} \Big( \langle {s_i s_j} \rangle_D -  \langle {s_i s_j} \rangle_M \Big)
\end{equation}

\begin{equation} \label{eq4}
\Delta b_{i} = \frac{1}{\eta} \Big( \langle {s_i } \rangle_D -  \langle {s_i } \rangle_M \Big) . 
\end{equation}
In equations \ref{eq3} and \ref{eq4}, the values inside  $\langle {*} \rangle$ represent expectation values over the data ($D$) and model ($M$) distributions.  The model would be perfectly trained if first ($\langle {s_i } \rangle$) and second ($\langle {s_i s_j } \rangle$) order moments were identical for both the data and model distribution.

To properly adjust the model parameters we need to calculate expectation values over the model itself.  Getting the ``true'' values would thus require a calculation for all $2^N$ possible states of the model, which is clearly intractable as the system size increases.  These particular calculations are where the use of quantum computation is ideal, and we see a potential for a speedup in our overall training algorithm.

The quantum devices produced by D-Wave Systems perform a quantum annealing algorithm.  In theory, this algorithm can leverage quantum effects to take an initial quantum system that is in a well-known ground state and transform this into a final Hamiltonian - one in which the system should still be in the ground state (assuming the annealing process was slow enough, as well as many other factors discussed elsewhere \cite{Farhi}).  The original use case of this algorithm lies in the fact that if you can properly map a computationally difficult problem of interest into this final Hamiltonian, then measuring the ground state of this final Hamiltonian should produce the optimal solution to the original problem.  However, this use case has been elusive at scale; as shown in the research of \cite{itay}, which focuses on fundamental limitations of quantum devices at finite temperatures.  Even taking some optimistic assumptions (such as perfect, instant thermalization), as the system (problem) size grows, the probability of measuring the optimal (ground) state of the system decreases exponentially.  Rather than returning the ground state solution, repeatedly measuring from such a device returns a Boltzmann distribution of energies.

While these results prove challenging for using such hardware for optimization, it presents an ideal opportunity for training BMs.  At a high level, instead of trying to calculate $\langle {s_i } \rangle_M$ and  $\langle {s_i s_j} \rangle_M$ directly, we can instead map our network onto the D-Wave quantum device.  By obtaining a finite number of samples from the hardware device, the goal is to generate better approximations of  $\langle {s_i } \rangle_M$ and  $\langle {s_i s_j} \rangle_M$ than classical heuristics. This method seems all the more natural as the form of the Hamiltonian $H$ of the D-Wave device is
 \begin{equation}  \label{eq5}
H(\textbf{S}) = -\sum_{S_i \in \textbf{S}} h_i S_i  - \sum_{S_i, S_j \in \textbf{S}} J_{ij} S_i S_j ,
\end{equation}
which is the same functional form as the BM energy in equation \ref{eq2}.  In this equation, \textbf{S} is the vector of qubit spin states, $h_i$ are the bias terms on each qubit, and $J_{ij}$ are the (anti)ferromagnetic couplings between the qubits.  By mapping the model parameters of a BM to the hardware parameters of a D-Wave device and making a set of measurements of the device, one can use these measurements to construct approximations of $\langle {s_i } \rangle_M$ and  $\langle {s_i s_j} \rangle_M$.  Advantages have been shown in using fully-connected BM on QC devices because using the methods of \cite{PhysRevX.7.041052}, the effective temperature of the device does not have to be taken into account.  Equation \ref{eq1} is a special case of the more general representation; rather than raising the exponential to $-E(\textbf{s})$, the more general expression raises its to $-E(\textbf{s})\beta$, where $\beta$ is the ``effective" temperature of the system (parameter related to temperature of the system).  If $\beta=1$ then we arrive at equation \ref{eq1}, but in general when using a quantum device one will not know the effective temperature beforehand, which can experience large fluctuations between measurements.  While this can be problematic for training RBMs using quantum annealers, and requires different techniques to estimate this parameter (\cite{Benedetti2016, 2016arXiv160606123D, arXiv:1510.06356}), fully connected BMs do not require these additional calculations for effective training (\cite{PhysRevX.7.041052}).

\begin{figure}[H]
	\centering
	\includegraphics[width=1\linewidth]{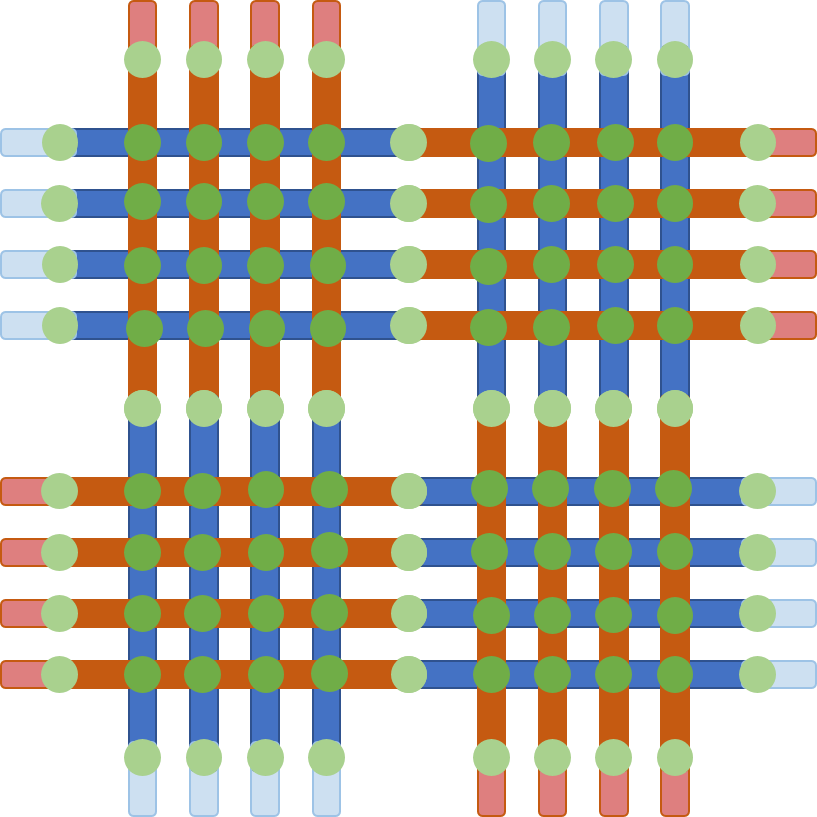}
	\caption{Four bipartite cells of a Chimera graph architecture showing how the cells interconnect. In each cell there are four horizontal and four vertical qubits, colored alternately blue and burnt orange. Within cells, where two qubits overlap they are coupled by means of a Josephson junction, indicated by green circles. Each qubit can be coupled to two addtional qubits from adjacent cells, also by means of Josephson junctions, indicated by light green circles.}
	\label{fig:chimera2}
\end{figure}

Though the structure of the BM graph to embed on the device is fully connected, we are in practice limited by the graph structure physically realized in the hardware. The adiabatic quantum device we used for this research was a D-Wave 2X, which has 1,152 qubits connected in a Chimera graph architecture consisting of 8 qubit cells arranged as $K_{4,4}$ bipartite graphs. The qubits within each cells are cross connected, and each cell is connected to four adjacent cells (with the exception of cells along the boundaries) as shown in Figure \ref{fig:chimera2}.  To properly map the BM energy function of (2) to the device, the graph minor-embedding problem must be solved; we need a hardware embedding which uses a chain of multiple physical qubits to realize a single logical qubit in the problem Hamiltonian of (5).  Using the same method as \cite{PhysRevX.7.041052}, we find embeddings using the embedding heuristic provided by D-Wave's API and resolve discrepancies of the qubit chains using majority vote (a post-processing step of the measurements).

\subsection{Quantum Boltzmann Machines for Forecasting Elections} \label{forecasting_elections}

The methodology outlined in section \ref{modeling_boltzmann_machines} lays out our approach for training fully-connected BM using a D-Wave quantum device.  This section will detail our procedure for implementing these networks to forecast elections.  In this research, we study the US Presidential election, and each binary unit in the BM represents a single US state.  The winner of a particular election simulation is determined by the candidate with the most electoral college votes.  Each US state has a particular number of electoral college votes to award to a candidate (2 + an integer which scales as a function of the state's population), and these votes are awarded entirely to one candidate (winner-take-all).  We assert that each sample returned from a fully-connected BM will in effect be a simulation of a US presidential election.  Each sample from the BM returns a binary vector, where each entry in the vector corresponds to the voting results of a particular US state.  These individual state voting results are mapped to a particular candidate/party (i.e., 1 = Democrat, 0 = Republican).  To determine the election simulation outcome, we weight each of these US state outcomes according to their net weight in the national vote (each state's electoral votes).  The winner of each simulation (sample) is determined by the sum of each party's overall national vote, which is calculated using the binary results (from sample) and national weight (electoral votes) for each state.

\begin{figure}[H]
	\centering
	\includegraphics[width=1\linewidth]{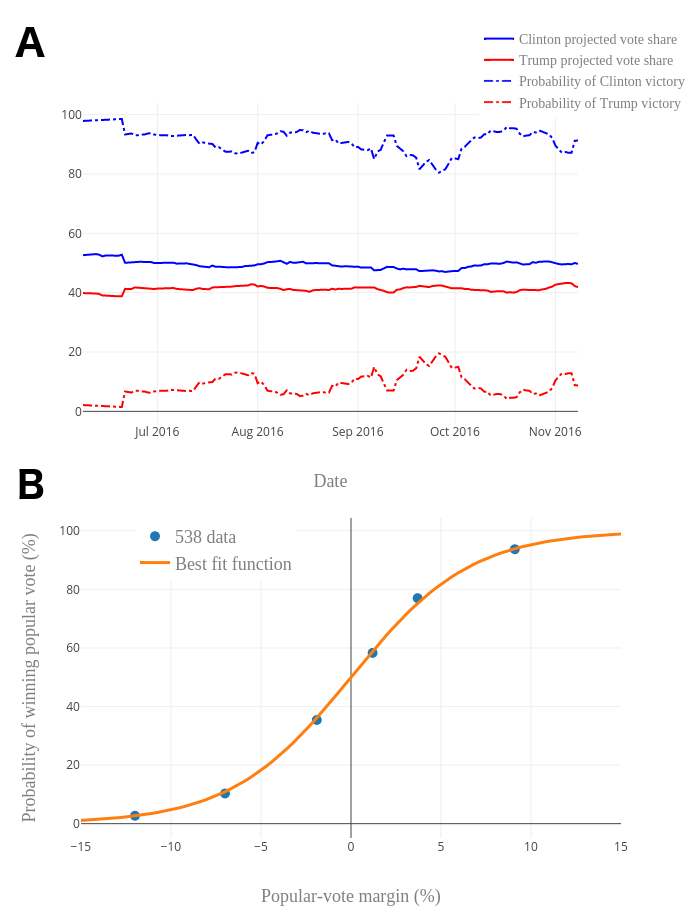}
	\caption{Model for interpreting projected vote share to probabilities.  A.  Plot of Maine's polling projections over time, where the solid lines are the time-averaged projected vote share for both candidates, and the dashed lines are the resulting probabilities of victory for each candidate, calculated using the best fit function shown in B.}
	\label{fig:Plot2}
\end{figure}

The goal is to train the BM which is being sampled from so that the first and second order moment terms of the model distribution approach those of the data distribution.  This training process has already been discussed in section \ref{modeling_boltzmann_machines}, and in this section we will expand on how we determined the first and second order moment terms for the data distribution of our election model.  The first order moment terms represent the probability that each state will vote for a particular candidate.  As an example, if we believe that there is a 80\% chance that the Democratic candidate wins Colorado, then the first order moment for the binary variable assigned to represent Colorado should be equal to $0.8$.  To determine all the first order moments for each state in our model, we use the current time-averaged polling results made publicly available on FiveThirtyEight (\cite{538model}).  We obtain a projected vote share for both candidates for each day that data is available (6 months before, and including, November 8th 2016).  These projected vote shares are then used as input to a sigmoidal model (same model used by FiveThirtyEight \cite{538sigmoid}) which rightly assumes that elections are stochastic, and that the result for each state/country follows a probabilistic rather than deterministic curve based on the popular-vote / projected vote share margin.  This method for converting a popular-vote margin to a probability of victory is shown for a particular state in Figure \ref{fig:Plot2}.

Given the underlying projected vote shares for each state and the best fit function shown in Figure \ref{fig:Plot2}B, calculating the first order moment terms for each state is straightforward.  Calculating the second order terms, the effective ``correlations" between states, is much more difficult.  These correlations express the likelihood that two states will end up with the same (or different) voting result in an election.  States that vote the same are more correlated (higher second order moment), and states that don't have a lower correlation (lower second order moment).  These correlations are influenced by a plethora of demographic (race, age, education), geographic, and additional factors.  Professional modelers (such as those at FiveThirtyEight) have complex methodologies for determining these correlations; however, a rigorous analysis of these correlations is outside the scope of this particular work.  We used data obtained publicly, which is sufficient to validate the general approach of our model.

To calculate the second order moment terms, we use one source of data and make two particular ansatz.  The data source we use is presidential election results from the last 11 US Presidential elections.  This data contains the date and results per state for each election.  Our first ansatz is that if we consider two states, these states should have higher correlations (second order terms) if they had voted similarly in previous elections.  This correlation is agnostic towards which candidate was voted for in each of these previous elections; the only important factor for the two states in question is if the vote was for the same candidate or a different one.  The second ansatz is that in terms of weighting previous election results, more recent elections are more relevant.  This means recent elections increase correlations between two states more than those that happened longer ago.  We assume a linear relationship between time and importance.  The raw correlations $\langle {s_i s_j} \rangle_{D_{raw}}$ between states $i$ and $j$ are calculated as follows:
\begin{equation}  \label{eq6}
 \langle {s_i s_j} \rangle_{D_{raw}} =  \frac{\sum_{n = 1:11} n \big( 2 i_n j_n - i_n - j_n + 1 \big)}{\sum_{n = 1:11} n}
 \end{equation}
wherein $n$ refers to a particular election year in the set [1968, 1972, ... , 2008, 2012] (higher $n$ is more recent) and $i_n$ and $j_n$ are the results for election $n$ for both respective states.  We then enforce a hard constraint that second order correlations should never contradict first order moments (which are calculated directly from current polling data).  This is accomplished by finally calculating the second order moments between states $i$ and $j$ as 
\begin{equation}  \label{eq7}
 \langle {s_i s_j} \rangle_{D} =  \langle {s_i s_j} \rangle_{D_{raw}} min(\langle{s_i} \rangle_{D}, \langle{s_j} \rangle_{D} ) .
 \end{equation}

We have now a methodology for mapping election forecasting models, specifically the 2016 US Presidential election, to BM by defining mathematical models for calculating both first and second order data distribution terms.  In the following section, we validate that this approach holds true for small, nonexistent countries and then attempt to simulate a ``real time" forecast for 2016 Presidential election using quantum-trained BMs.

\subsection{Caveats and limitations} \label{caveats}
In section \ref{modeling_boltzmann_machines} and \ref{forecasting_elections}, we reviewed the methodology for training fully-connected BM with a D-Wave machine as well as describe our approach for mapping election forecast models to the (to be trained) BM.  While this work uses this approach as described, a few caveats and limitations deserve some additional attention here.

\subsubsection{Hardware constraints} \label{hardware_constraints}
The hardware size limitations of the D-Wave 2X does not allow us to fully embed a 50 state model as well as the DC province, which are the 51 fundamental voting blocks for the US Presidential election.  Using the virtual full-yield Chimera capability offered by D-Wave, which uses a combination of the quantum device hardware in tandem with classical software for simulating missing qubits and couplers, we were able to embed 49 states and omitted DC and Maryland.  These were omitted because they were ranked as the most ``definite" by model standards (both were approaching 100\% likelihood to vote Democrat), as well as geographically adjacent.

\subsubsection{Assert that all states are winner-take-all} \label{winner_take_all}
While the US Presidential election is winner-take-all at the state level, two states are exceptions to this rule: Maine and Nebraska.  Instead of winner-take-all, these states award delegates by district.  To simplify the model and fit within the hardware constraints, we treat these states as winner-take-all regardless.  This decision was made for three reasons.  First, the primary purpose of this paper is to validate the overall election methodology for modeling such elections using QC-trained neural networks; such state specific rules fall outside this scope of this work.  Second, these states have small weight (electoral college votes) in the broader election, so treating them as winner-take-all has a reduced effect compared to a much larger state under the same voting system.  Third, in the future we could treat the provinces as individual states themselves, each awarding electoral college votes with a winner-take-all system.  However, due to our limitation already expressed in the previous issue, this experiment will be left to future studies on a larger quantum device.

\subsubsection{Inability to model national errors in same model} \label{national_errors}
The strength in models with correlations as described here is simple; they can account for a form of error that is inaccessible to independent models.  However, there are also two other primary types of error that we would want our final model to consider: national and state-specific errors.   Both of these error arise from the fact that polling is never perfect; there are always voting blocks that are under or over-represented based on the types of people that are both polled and respond to the poll.  National error arises from the fact that all states could have systematically missed a particular type of voting block in a similar, characteristic manner.  This leads to errors that affect each state in a similar way.  State-specific error is the same concept, but on a state-by-state level.  The latter (as discussed in future results section) can be addressed naturally by the nature of the QC-training algorithm; however the former cannot.  Since we wish to emulate the best possible model, incorporating state-specific, state, and national errors, we choose to create a meta-model which aggregates results from several different models build on the assumption of different national error.  In our case, we take 25 equally-spaced samples from a t-distribution with 10 degrees of freedom; this is the same distribution and degrees of freedom used for national and state-specific error used by FiveThirtyEight (\cite{538model}).  These points define the national errors we use to train 25 different models.  For instance, one national model may have a national error favoring Clinton by 1\% point, while another might favor Trump by 1\%.  These national errors are ultimately incorporated into the first order moment terms for each state, leading to models which are slightly biased towards either candidate to a relative degree.  The average of these 25 models is calculated after simulating each model independently, and weighing the results of each by the probability of occurrence for each national error.

\subsubsection{Limited time windows of D-Wave access} \label{limited_time}
In a production environment, it would be ideal to produce updates to forecasts daily (or sometimes several times a day) for particularly high-profile elections.  These updates occur as new polls come in, changing the particular predictions for each state, and thus ultimately the national results. Applying our proposed methodology could assuredly be used for these purposes, but a limiting factor for simulating this daily forecast over 6 months is access to the D-Wave quantum device.  Due to limited access time to run experiments on the D-Wave device, and that we have to simulate multiple error models (as explained in section \ref{national_errors}), we choose to only model every 2 weeks of data rather than daily.  This allowed us to generate an appropriate number of simulations for these days across all national error models.

\section{Results} \label{results}

\subsection{Effect of Chain Length} \label{chain_length}

As mentioned in section \ref{modeling_boltzmann_machines}, using a well posed Hamiltonian and the right environmental variables, an AQC should theoretically be capable of finding the ground state of the Hamiltonian.  In practice, thermal fluctuations, environmental interactions, insufficiently short annealing times, and a plethora of other physical and engineering challenges result in a low probability of measuring the ground state, but instead some other low energy (potentially near-optimal) state.  This is especially true for larger Hamiltonian systems, as shown in \cite{itay}; for finite-temperature AQCs, as the system size increases, the probability of measuring a non-optimal low energy state approaches 1.  In contrast, if we wish to use an AQC as a sampling engine for sampling from BMs, we can potentially face a different set of obstacles when using small embeddings (system sizes).  In terms of using an AQC for machine learning purposes, returning a distribution of low energy solutions rather than the optimal configuration drives the learning process, as the first and second order statistics of these measurements determines the update terms for the model.   At small physical embedding size, the probability of measuring the optimal state increases significantly, and at very concise embedding sizes the probability of measuring the ground state energy can approach 1.  This behavior is that of a Hopfield network, which is a BM at $T = 0$.  Unlike a BM, the Hopfield network can only return ground state energy solutions.  This would imply that our training algorithm as described in section \ref{methodology} would not work for such a system.  Each time model updates in equations \ref{eq3} and \ref{eq4} are made, the energy function of equation \ref{eq2} changes as well.  This new energy function would lead to new ground state solutions, which in turn could have completely different model parameters.  While training a BM leads to model updates ``smoothly" guiding the model parameters ($\langle {s_i } \rangle_M$, $\langle {s_i s_j} \rangle_M$) towards the data distribution ($\langle {s_i } \rangle_D$, $\langle {s_i s_j} \rangle_D$), slight changes in the model parameters of a Hopfield network can completely change the ground state solutions, leading to chaotic model parameter updates.

\begin{figure}[H]
	\centering
	\includegraphics[width=1\linewidth]{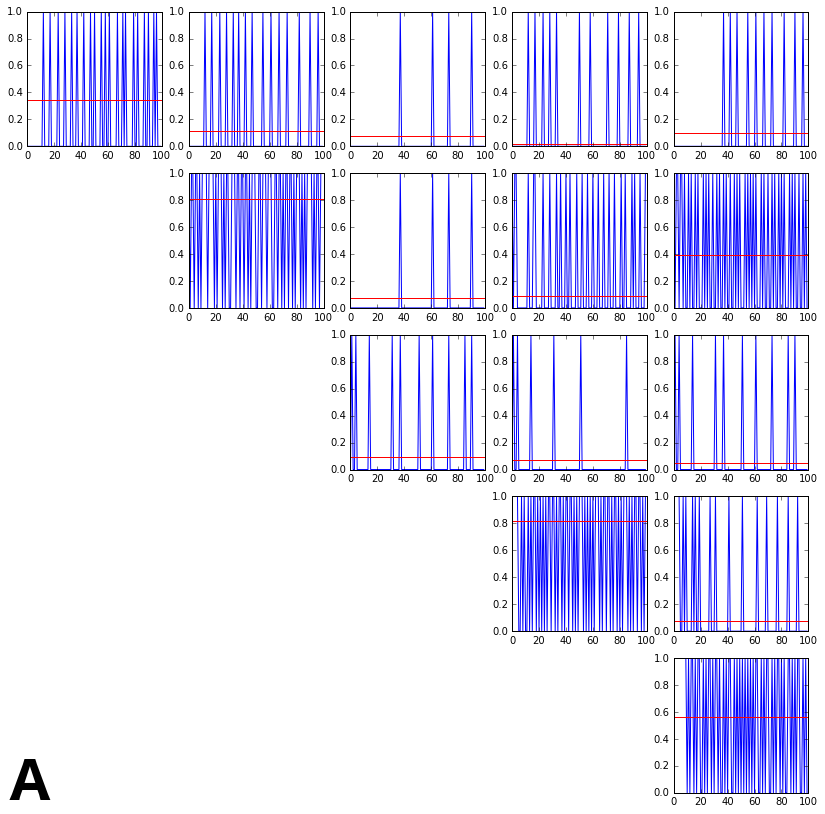}
	\includegraphics[width=1\linewidth]{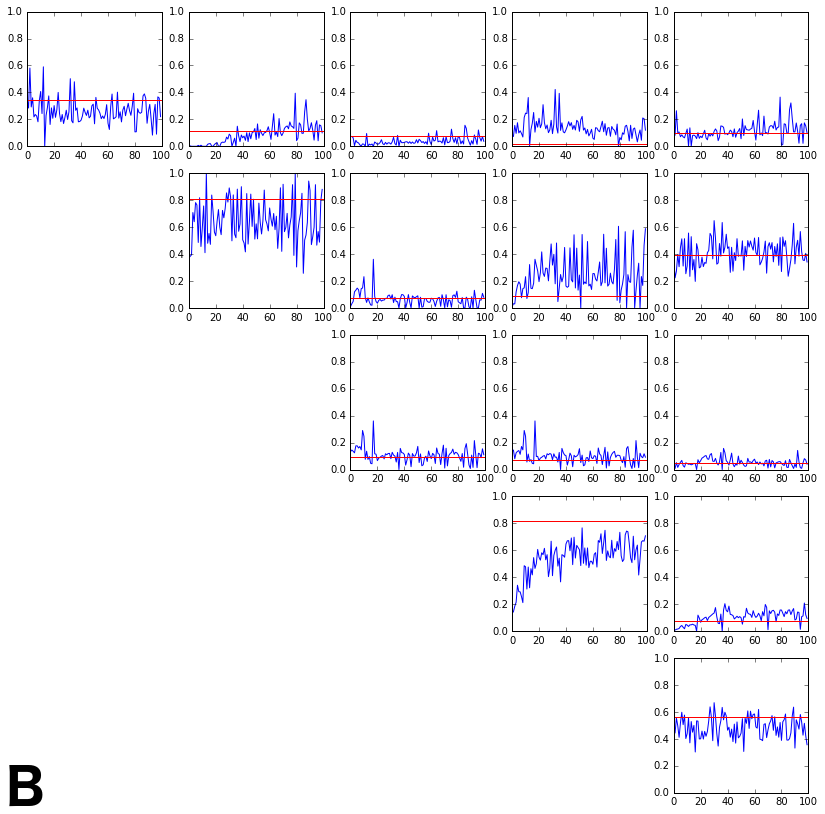}
	\caption{Training results for arbitrary Boltzmann machines realized on D-Wave device using (A) 1x and (B) 2x embedding qubit embedding chains.  In each subplot, the horizontal red lines are the respective target values.}
	\label{fig:5hardware1x}
\end{figure}

One potential way to mitigate these effects is to deliberately increase the size of the qubit chains for embedding the problem.  For optimization purposes, the goal would be to find the minimum chain length for embedding the problem Hamiltonian onto the physical device.  By keeping the embedding chains minimal, the system size is as small as possible which increases the chance of measuring an optimal ground state.  The opposite should be true as well: the more we increase the chain lengths for embedding the logical qubits onto the hardware, the more low energy states become available to system, increasing the probability that the system will transition away from the ground state during the annealing process.  This should enable one to properly train BM for any number of nodes, given that the qubit chain lengths are long enough.  By validating this assertion, we can argue that our approach here for using AQC for realizing BMs for election modeling could be applied to any sized system, as well as validate that our particular experiments are in a regime where proper learning is possible.

To test this hypothesis, we performed experiments with fully connected graphs of size 5 through 9, embedded with three different embeddings of various chain lengths, and studied how well we could train the systems to reproduce activation probability distributions defined by graphs with arbitrary first and second order terms.  For each run, the activation ($h_i$) and correlation ($J_{i,j}$) probabilities were selected randomly such that for node $i$ the activation probability $h_i \in (0, 1)$ and for two nodes $i$ and $j$, the correlation probability $J_{i,j} = c_{i,j}h_ih_j$ where $c_{i,j}$ is the correlation strength and $c_{i,j} \in (0, 1)$.

Three embeddings were used for each graph: a maximally concise embedding, an embedding derived from a maximally concise graph of twice as many nodes (denoted by ``2x"), and an embedding derived from a maximally concise graph of three times as many nodes (denoted by ``3x"). The decision to approach the problem in this way was done because the D-Wave API has been set up for optimization problems, and as such the hardware embedding functions in general attempt to return maximally concise embeddings. The 2x and 3x embeddings returned from the API were for graphs of 2x and 3x the size of the problem graph, so they were reduced to the correct size by joining the physical qubits representing pairs (in the case of 2x) and triples (in the case of 3x) of logical qubits (usually represented by chains of physical qubits) into single logical qubits of chains of physical qubits 2x and 3x times as long as in the original embedding.  An example of training by the shortest (1x) and medium (2x) chain lengths are shown in Figure \ref{fig:5hardware1x}.

For all the subgraphs of Figure \ref{fig:5hardware1x}, the x-axis of each graph is the number of completed iterations in the training algorithm while the y-axis is the activation probability when sampling the graph multiple times.  The graphs on the diagonal are single node activation probabilities(first order moments) and the off-diagonal graphs are the two node correlations (second order moments).  In Figure \ref{fig:5hardware1x}A, the activation probabilities fail to converge to the desired values, indicating that the qubit chains are not allowing sufficient degrees of freedom for the system to model a Boltzmann machine.  However, using the same network but with the 2x embedding qubit chains, the network was able to converge over time towards the target first and second order moment values.  In Table \ref{table:chainRMSE}, we show the root mean squared error (RMSE) for training iterations 191-200 for different QC-trained networks at different embedding chain lengths.

\begin{table}[H]
	\centering
	\begin{tabular}{c | c | r}
		Nodes & Chain & RMSE \\
		\hline
		5 & 1x & 0.437 $\pm$ 0.072 \\
		5 & 2x & 0.106 $\pm$ 0.036 \\
		5 & 3x & 0.060 $\pm$ 0.038 \\
		9 & 1x & 0.149 $\pm$ 0.101 \\
		9 & 2x & 0.038 $\pm$ 0.028 \\
		9 & 3x & 0.045 $\pm$ 0.034 \\
	\end{tabular}
	\caption{RMSE for QC-trained networks at different embedding chain lengths.  As the networks grow larger, the chain length differences grows more negligible as chains are naturally getting longer to satisfy the embedding.}
	\label{table:chainRMSE}
\end{table} 

Given the current D-Wave qubit connectivity graph, as the problem size grows larger, the average embedding chain length similarly grows.  As most studies  embed as large a problem as possible onto the device, this has naturally led to longer chain lengths in previous research.  As future hardware improvements are made and shorter qubit chains are feasible (through increased connectivity), it may become important to validate that the individual logical qubits are properly learning the respective target terms.  The lengthening technique shown here could provide a simple but efficient tool for ensuring Boltzmann-like behavior for all nodes in the logical graph without having to perturb any of the individual energy scalings.

\subsection{Modeling the Presidential Election} \label{presidential_results}

The primary experiment we conducted was to attempt to simulate a ``real-time election model forecast" using QC-trained Boltzmann machines.  Starting on  the date 2016-06-08 and continuing until election day 2016-11-08, we trained multiple fully connected Boltzmann machines using the D-Wave adiabatic device.

\begin{figure}[H]
	\centering
	\includegraphics[width=\linewidth]{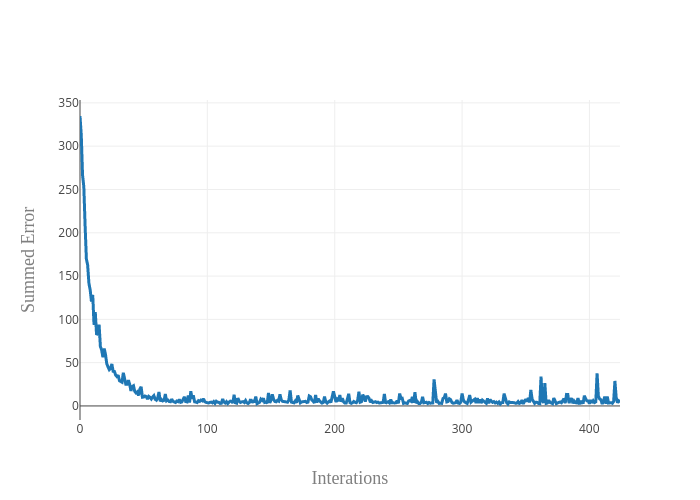}
	\caption{Summed error as a function of training iterations for one national error model.  The small spikes of error occur deep into the training process are simply an artifact of the updates in the first order moments that happen at 2 week (25 iteration) intervals.}
	\label{fig:training-error}
\end{figure}

\begin{figure*}
\centering
  \includegraphics[width=1\textwidth]{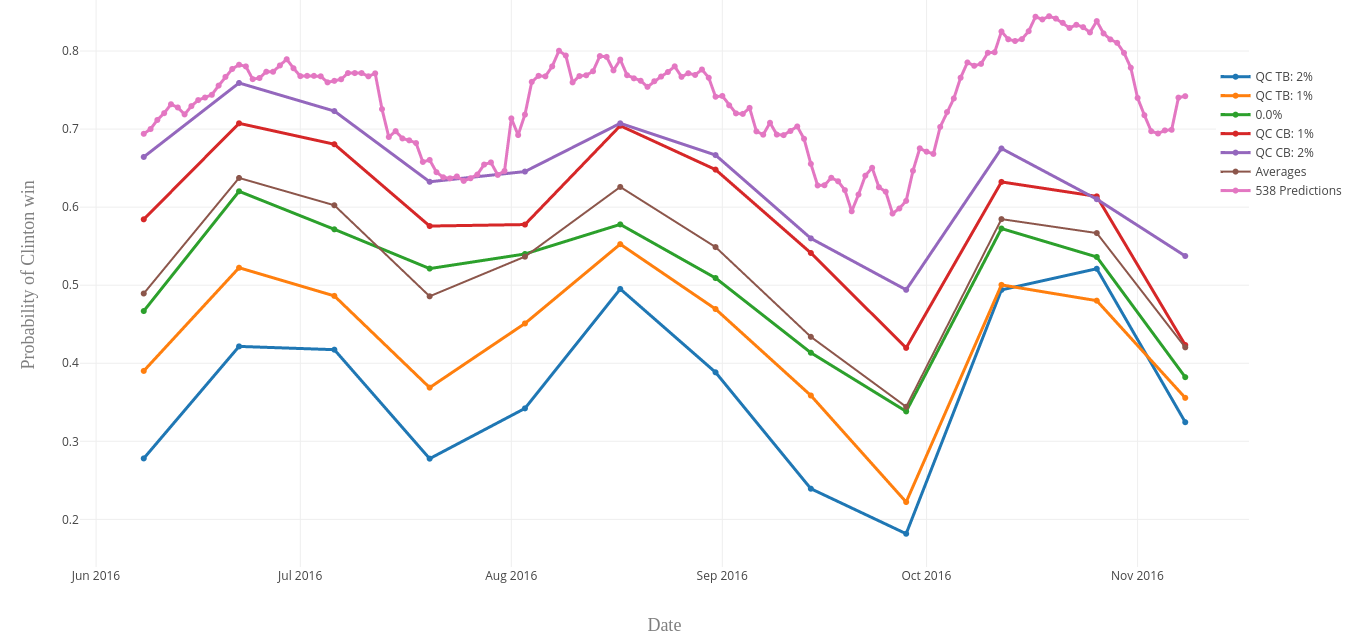}
  \caption{Comparing 2016 Presidential election forecasting results from QC-trained methodology to those of FiveThirtyEight.  QC-trained networks each had a national bias towards Clinton (CB), Trump (TB), or neither candidate.}
  \label{fig:election_result}
\end{figure*}

As mentioned in section \ref{national_errors} and \ref{limited_time}, due to limitations we retrained the network every two weeks rather than daily, and used 25 different networks to model different national errors (derived from a t-distribution with 10 degrees of freedom).  The networks starting on 2016-06-08 were initialized with small, random coefficients and then subsequently trained for 150 iterations each.  Then, at each 2 week interval, the first order moment terms were updated and trained for an additional 25 iterations.  The changes to the first order moments were small, so fewer training iterations were necessary to converge to a stable summed error (sum of squared first and second moment errors) across the networks.  This led to 400 total training iterations per national error (150 for 2016-06-08 + 25*10 for the next ten two-week updates).  An example of the training error for a particular national error model is shown in Figure \ref{fig:training-error}. 

Knowing from section \ref{chain_length} that our qubit chains are sufficiently long enough to learn properly, the training error results of Figure \ref{fig:training-error} are to be expected.  Similar plots were observed across all national error models, as this translates into nothing more than scaling the first order moment terms.  We can then take the samples from these networks at different iterations as our election forecasting simulation results.  We choose to take samples for the last 10 iterations of each forecasting date (this would be iterations 141-150 for 2016-06-08 and 16-25 for the next 10 forecasting dates).  This allows us to sample from the network once it has reached a general steady-state in terms of summed training error.  As discussed in section \ref{forecasting_elections}, each logical qubit is mapped to a particular state and each sample is equivalent to an election forecast.  To determine which candidate ``won" a particular sample, we simply map the qubit results back to the particular state it represents, and add each state's number of electoral votes to the candidate that state voted for in the sample.  Since the democratic candidate was the heavy favorite in most election models, we choose to express our forecasting results in terms of the probability of a Clinton victory.  In this way, each sample results in a particular candidate winning (270 electoral votes or more) or losing (we combined ties into this category for simplicity, although an exact tie is very unlikely).  For our experiments, we took 1,000 samples from the D-Wave device at every iteration for each national error model.  This gave us 10,000 samples for each national error model for each forecasting date (10 training iterations, 1,000 samples per training iteration).  The probability of a Clinton victory for each national error model was simply the sum of the individual samples which were won by Clinton divided by the number of total samples (10,000 in our case, per national error model and time step).  Finally, to get an average election forecast as a function of time (shown in Figure \ref{fig:election_result}), we calculated the weighted arithmetic mean across all national error models for each forecasting date.  The weights for each national error were defined as the t-distribution probability density function of each national error (t-distribution with 10 degrees of freedom).

As evidenced in Figure \ref{fig:election_result}, the QC-trained network results followed trends similar to the trends of the professional FiveThirtyEight forecasts.  The overall probabilities of the different national error networks also follows naturally; networks that had a national error in favor of Clinton increased the probability of a Clinton victory, and networks with a national error in favor of Trump decreased the probability of a Clinton victory.  The largest apparent difference between the QC-trained models was the overall probability of a Clinton victory.  While the Average result line of the QC-trained networks follows a very similar pattern to the predictions of FiveThirtyEight, the QC-trained results are almost uniformly ~20\% lower.  This result in no way says the quantum methodology is ``better", but rather highlights the differences in the overall approach.  It is likely that these results are mostly dependent on the underlying differences in how we calculated our second order moments terms between the states.  An interesting future study would be to replicate the quantum-training protocol described here but using second order moments driven by demographic data of the individual state inhabitants.

An important factor forecasters also desire from a forecasting model is to know which states are the most important for predicting a particular outcome.  A straightforward approach is to generate a vector for each state (1 = state voted Democrat, 0 = state voted Republican) and a similar vector for the outcome (1 = Democratic victory, 0 = Republican victory) of each simulation for the date November 8, 2016.  Then, we can calculate the Pearson correlation coefficient between the two vectors and take the absolute value of these correlations.  Table 2 shows the 10 states with the highest and lowest correlation coefficients.  As expected, states that leaned heavily Democratic or Republican had very low correlation coefficients; regardless of the outcome of the election, states like Illinois and Nebraska were virtual locks for the Democratic and Republican candidates, respectively.  Similarly, the states with the highest correlation coefficients contained many of the most contested states in the election.  FiveThirtyEight's forecasts have a similar ``tipping-point chance" metric which they define as ``the probability that a state will provide the decisive vote in the Electoral College" (\cite{538model}).  On election day, 7 out of the 10 states they ranked as the highest tipping-point chance states were similarly in the list of 10 most correlated states in Table \ref{table:stateFeatureImportances} (the differences: FiveThirtyEight included Virginia, Minnesota, and Wisconsin, while ours included New Hampshire, Iowa, and Arizona).

\begin{table}[h]
	\centering
	\begin{tabular}{c | c}
		State & Correlation coefficients \\
		\hline
		Ohio & 0.204 \\
		Florida & 0.163 \\
		Nevada & 0.178 \\
		New Hampshire & 0.167 \\
		Pennsylvania & 0.155 \\
		Iowa & 0.152 \\
		Michigan & 0.145 \\
		North Carolina & 0.137\\
		Colorado & 0.130\\
		Arizona & 0.127\\\cline{1-2}
		Illinois & 0.002\\
		Nebraska & 0.004\\
		Alabama & 0.005\\
		Oklahoma & 0.006\\
		California & 0.008\\
		West Virginia & 0.008\\
		Delaware & 0.008\\
		Oregon & 0.009\\
		Idaho & 0.015\\
		Arkansas & 0.016\\
	\end{tabular}
	\caption{Pearson correlation coefficients for the 10 states most (top) and least (bottom) correlated with the election forecasting results.}
	\label{table:stateFeatureImportances}
\end{table} 

\begin{figure}[H]
	\centering
	\includegraphics[width=\linewidth]{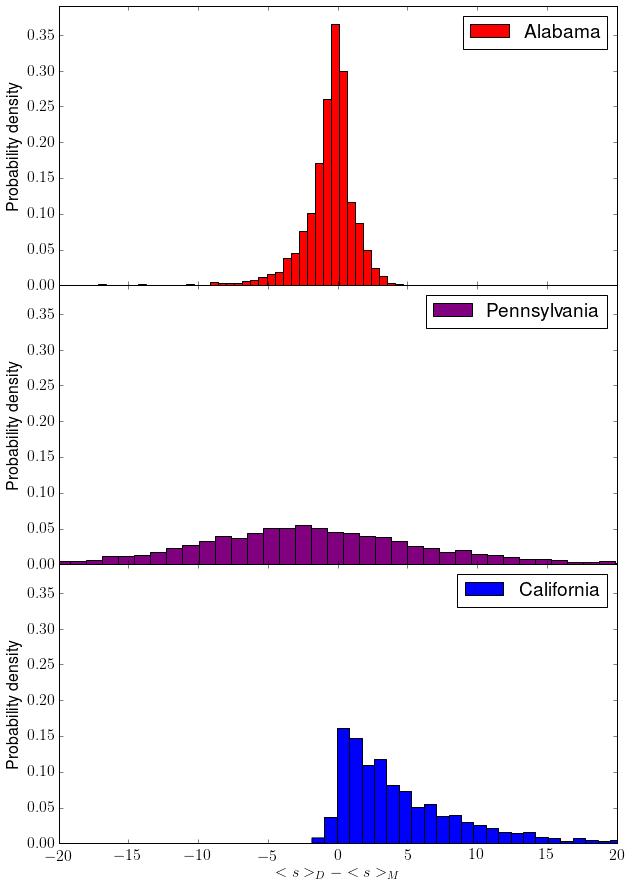}
	\caption{Example distributions of state-specific error for states leaning heavily Republican (top), Democratic (bottom), or swing states (middle).}
	\label{fig:state-error}
\end{figure}

Finally, we take into consideration the individual state errors observed in our QC-trained models.  As mentioned previously, modelers (such as FiveThirtyEight) will apply some degree of noise for individual states, such as adding in state-specific error from sampling.  It would be useful to know how the natural sampling of the quantum device during training lends itself to state-specific error.  For each iteration that we used for determining the national averages, we calculated the difference between the target $\langle {s } \rangle_D$ and current model output $\langle {s } \rangle_M$.  If this difference is negative, this would be a state-specific error in favor of the Democratic candidate, and vice versa a positive value translates to error benefiting the Republican candidate.  By taking all these errors per state, we can form different state-specific error distributions per state.  These distributions vary considerably, depending on the underlying target $\langle {s } \rangle_D$ value, as evidenced in Figure \ref{fig:state-error}.

At the extremes, we see that the error distributions of states leaning heavily Democratic or Republican are asymmetrical.  This occurs naturally, due to $\langle {s } \rangle_M$ being bound between 0 and 1.  If $\langle {s } \rangle_D \approx 0$ (state leaning heavily Republican), all error will be biased in the negative direction; similarly, states with $\langle {s } \rangle_D \approx 1$ (state leaning heavily Democrat) will have positively-biased error distributions.  For swing states, we see a much more uniform spread of error, which shows that in the absence of bounds ($\langle {s } \rangle_D \approx$ 0 or 1), the error tends to be equally distributed.  

One interesting finding was that heavily-learning Democratic states seemed to have longer error distribution tails compared to the heavily-leaning Republican states.  As seen in Figure \ref{fig:state-error}, while almost all the probability mass of Alabama's error distribution is contained within the range -5 to 5, a substantial amount of California's error distribution falls outside these bounds.  This phenomena can introduce an amount of bias in favor of one particular candidate.  One potential mitigation technique for dealing with this issue is taking the average of multiple gauges (\cite{arXiv:1510.06356}), some of which could ``flip" the measurement value (flip Republican = 1, Democrat = 0).  Additionally, some interesting new techniques using ``shimming" (\cite{shimming}) have been shown to reduce overall qubit error.  In future work, it would be an interesting topic to explore the evolution of individual logical qubit error distributions in QC-trained Boltzmann machines by using shimming techniques (reducing error) or introducing random noise (increasing error) on a per-qubit basis.

\section{Conclusions}

In this work, we have showed an initial implementation of QC-trained Boltzmann machines, which can be employed for the difficult task of sampling from correlated systems, an essential problem at the heart of many applications such as election forecast modeling.  We validated that this approach successfully learned various data distributions based on state polling results during the 2016 US Presidential campaign, and these QC-trained models generated forecasts that had similar structural properties and outcomes compared to a best in class election modeling group.  While quantum computers and samplers are an emerging technology, we believe this application area could be of near-term interest.  This methodology could be an interesting technique to bring to the broader conversation of modeling in future election forecasts.

\bibliographystyle{h-physrev}
\bibliography{ElectionModeling}

%\end{multicols}

\end{document}